\documentclass[11pt,superscriptaddress,aps,prd]{revtex4}

\everymath{\displaystyle}
\usepackage{graphicx,amsmath,amssymb,amsfonts,amstext,latexsym,stmaryrd,amscd,txfonts,pxfonts,hyperref}
\hypersetup{
    colorlinks=true,
    linkcolor=blue,
    filecolor=magenta,      
    urlcolor=black,
		citecolor=red,
		}

\begin{document}
\title{On the total energy conservation of the Alcubierre spacetime}

\author{F. L. Carneiro}
\email{fernandolessa45@gmail.com}
\affiliation{Instituto de F\'isica, Universidade de Bras\'ilia, 70910-900, Bras\'ilia, DF, Brazil}

\author{S. C. Ulhoa}
\email{sc.ulhoa@gmail.com}
\affiliation{Instituto de F\'isica, Universidade de Bras\'ilia, 70910-900, Bras\'ilia, DF, Brazil}
\affiliation{International Center of Physics, Instituto de F\'{\i}sica, Universidade de
Bras\'{\i}lia, 70.910-900, Brasilia, DF, Brazil}
\affiliation{Canadian Quantum Research Center,\\ 
204-3002 32 Ave Vernon, BC V1T 2L7  Canada}

\author{J. W. Maluf}
\email{jwmaluf@gmail.com}
\affiliation{Instituto de F\'isica, Universidade de Bras\'ilia, 70910-900, Bras\'ilia, DF, Brazil}

\author{J. F. da Rocha-Neto}
\email{rocha@fis.unb.br}
\affiliation{Instituto de F\'isica, Universidade de Bras\'ilia, 70910-900, Bras\'ilia, DF, Brazil}

\begin{abstract}

In this article, we consider the Alcubierre spacetime, such a spacetime describes a ``bubble'' that propagates with arbitrary global velocity.  This setting allows movement at a speed greater than that of light.
There are some known problems with this metric, e.g., the source's negative energy and the violation of the source's energy conservation when the bubble accelerates. 
We address these two issues within the realm of the Teleparallel Equivalent of General Relativity (TEGR). The energy conservation problem can be solved when considering the energy of the gravitational field itself. The total energy of the spacetime, gravitational plus source, is conserved even in accelerated motion.
We explicitly show the dependence of energy and gravitational energy flux on the frame of reference, one adapted to a static observer and the other to a free-falling one in the same coordinate system.
Addressing the problem of energy negativity of the source, we find that a static observer measures positive energy of the source, while an Eulerian observer measures a negative one. Thus, we surmise that negative energy may be a reference problem.

\end{abstract}

\maketitle

\date{\today}

\section{Introduction}

Alcubierre spacetime was first introduced by Alcubierre in 1994 \cite{alcubierre1994warp} by what is now called ``metric engineering'' \cite{white2003discussion}. Some solutions to Einstein equations were obtained by establishing symmetries and then solving the equations themselves, including the first known solution, that of Schwarzschild.  Alcubierre spacetime was not obtained by integration of a energy-momentum tensor of a source but was constructed by geometric considerations and requirements.  Alcubierre's main objective in establishing such a solution was to obtain a geometry that would allow a spacecraft to travel at arbitrarily high speeds.  It should be noted that special relativity prohibits massive objects from reaching the speed of light thus from this point of view they cannot exceed this speed.  Within the realm of General Relativity (GR) there is also a similar prohibition, that is, local travelers cannot be faster than the speed of light.  Thus the Alcubierre metric seeks to describe theoretically how travelers can have arbitrary global speeds.  Although there are certain features considered undesirable in this spacetime, such as the violation of energy conditions and the presence of so-called closed timelike curves (CTCs), it determines a class of valid solutions to Einstein equations.  As we will see, some of these features can be reinterpreted to increase the scope of what would be physically acceptable solutions.  Particularly the violation of energy conservation can be avoided when taking into account gravitational energy.

The motion of test particles in Alcubierre spacetime is attained by constructing a non-trivial geometric configuration in the form of a bubble, where inside the bubble the spacetime is conformally flat and outside is Minkowski spacetime. The bubble may propagate at any global velocity $v_{s}$ and the proper time within the bubble is the same as the proper time of an observer at spatial infinity (for a definition see \cite{ashtekar1992spatial}), e.g., a static observer on Earth.
This important feature allows an observer to travel to a distant star and return without problems related to time dilation. Albeit time dilatation is not the only physiological problem that the traveler inside the bubble must deal with. The bubble's interaction with external particles results in their acceleration and radiation emission \cite{mcmonigal2012alcubierre}. Alcubierre spacetime, geodesically, consists of a congruence expansion converging in front (between the bubble and the destination) and diverging in the rear (between the bubble and the origin), but the divergent congruence, or ``geodesic expansion'', is not a necessary feature \cite{natario2002warp}.

Ever since the first appearance of Alcubierre spacetime, many authors reviewed this topic, including Alcubierre himself \cite{alcubierre2017warp}. Clark and collaborators studied the effect of the spacetime on null geodesics \cite{clark1999null}. McMonigal and collaborators treated the interaction of the bubble with massive and null particles \cite{mcmonigal2012alcubierre}. Addressing some causality problems in the Alcubierre spacetime, Krasnikov proposed modifications to the original spacetime, describing what is now called ``Krasnikov tube'' \cite{krasnikov1998hyperfast,everett1997superluminal,gravel2004simple}.
The source of the gravitational field was recently addressed by Santos-Pereira and collaborators who considered dust as the source of the gravitational field, obtaining a zero energy density in case of non-violation of the strong energy condition. Interestingly, in the latter case, an equation was obtained and that resembles the Burgers equation, which describes shock waves moving in a fluid. Latter, a perfect fluid and a charged dust were also considered as a source \cite{santos2021fluid,santos2021charged}.

There are many known physical limitations of the Alcubierre spacetime \cite{lobo2004fundamental,everett1996warp}, with several debates regarding the violation of energy conditions \cite{lobo2003weak}. Recently Bobrick and Martire presented an algorithm for constructing physical ``warp drives'' that avoids some of the problems present in Alcubierre spacetime \cite{bobrick2021introducing}. It was noted in this former paper that the Alcubierre spacetime has a problem in real situations. A real spaceship is not expected to always have a constant global velocity $v_{s}$, it is necessary for the ship to accelerate and decelerate, which implies that $v_{s}=v_{s}(t)$. When there is a temporal dependence of velocity, the $00$ component of the source's energy-momentum tensor also varies with $t$, so it was argued that an accelerating spacecraft violates the energy conservation. 

The main objective of the present paper is to address the energy conservation of the original Alcubierre spacetime. Although the energy of the source of Alcubierre spacetime varies with time in an accelerated motion, we find that the gravitational energy also varies in the same way, as a consequence the energy of the source and the gravitational energy exactly cancel each other at long distances from the bubble. Thus, the total energy of the spacetime is conserved even in accelerated motion. We also obtain positive energy for the source measured by a static observer and negative energy measured by a free-falling one. Therefore, we point out the possibility that the negative energy of the source in Alcubierre spacetime might be a problem of the frame. We also investigate the gravitational energy flux measured by static and free-falling observers, in the same coordinate system, and find a non-zero flux measured by both observers. 

In order to evaluate the gravitational energy and gravitational energy flux of the spacetime, we use the Teleparallel Equivalent of General Relativity (TEGR) which, although it has the same dynamics as GR, allows the definition of an adequate energy-momentum four-vector. In section \ref{terg} we review some aspects of the description of gravity in terms of the TEGR. In section \ref{review}, we briefly review Alcubierre spacetime and some of its properties, while establishing the coordinates that are used in this article. In section \ref{static}, we introduce a set of tetrads, associated with the Alcubierre metric and adapted to a static observer, and evaluate the $(0)$ component for the energy-momentum four-vector measured by this observer. In section \ref{freefall}, we apply a Lorentz transformation to the tetrad of the static observer and evaluate the $(0)$ component for the energy-momentum four-vector measured by the free-falling observer and the gravitational energy flux. In section \ref{vt} we address the case of variable velocity. Finally, in section \ref{conclusions} we present our conclusions regarding the obtained results and conjecture the implication of the results.

In this paper we have established the following notation. The Latin letters at begging of the alphabet $a$, $b$, $\ldots$ denote $SO(3,1)$ indices and local coordinates are indicated by $(0)$, $(1)$, $\ldots$. The Greek letters $\mu$, $\nu$, $\ldots$ denote the spacetime indices and spacetime coordinates are indicated by $0$, $1$, $\ldots$. Both run from $0$ to $3$. Time is indicated by $(0)$ and $0$ and space by $(i)$ and $i$, where $i=1,2,3$. The flat Minkowski tensor $\eta_{ab}$ raises and lowers $SO(3,1)$ indices; the metric tensor $g_{\mu\nu}$ raises and lowers the spacetime indices. The geometrized unit system is used, i.e., $G=c=1$.

\section{Teleparallel Equivalent of General Relativity (TEGR)}\label{terg}

The TEGR is a geometric description of gravity that yields field equations equivalent to GR. In GR, physical quantities are established in terms of pseudo-tensors \cite{landau2013course}, these are coordinate dependent. It is well known that there are difficulties in trying to establish a true energy-momentum tensor for the gravitational field in GR. The main objective of this paper is to analyze the conservation of the total energy, gravitational plus source (matter-radiation fields), of Alcubierre spacetime. In order to obtain consistent results, we need a properly established definition of gravitational energy that does not depend on coordinates. The TEGR, which uses a tetrad description of gravity, allows the definition of a proper energy-momentum tensor for the gravitational field and, as a consequence, an energy-momentum four-vector.
Thus, in this section, we briefly review some aspects of the TEGR that are used in this paper and present the expressions for energy due to gravitational field plus matter and the gravitational energy flux. A complete review of TEGR can be found in Ref. \cite{maluf2013teleparallel}.

Let us consider the spacetime as an orientable manifold in which, at any event $x^{\mu}$, we may establish a local frame in the tangent spacetime parametrized by $x^{a}$. The projection of spacetime quantities may be achieved through a tetrad field $e^{a}\,_{\mu}$, e.g., for a vector $V^{\mu}$ we have $V^{a}=e^{a}\,_{\mu}V^{\mu}$. Whereas in the absence of gravity the spacetime and the tangent space coincide and the quantities $dx^a$ are holonomic, in the presence of gravity, and non-trivial geometries, the quantities $dx^{a}$ are non-holonomic. When considering the world line of an observer $x^{\mu}(\tau)$, where $\tau$ is the proper time, the components $e_{(0)}\,^{\mu}$ are always tangent to the observer's worldline. Thus we identify $e_{(0)}\,^{\mu}$ with the observer's four-velocity $u^{\mu}$ in its rest frame.

In the TEGR, the gravitational fundamental variables are the tetrads $e^{a}\,_{\mu}$. The metric tensor can be obtained from the tetrads as $g_{\mu\nu}=e^{a}\,_{\mu}e_{a\nu}$. While the metric tensor has ten independent components, the tetrads have sixteen. 
Thus, any quantity that is a function of the metric is also a function of the tetrads, but the inverse is not true. By choosing the six additional components, a reference system is established. The theory is constructed on the geometry of absolute parallelism. Therefore,
\begin{equation}\label{eq1}
\nabla_{\mu}e^{a}\,_{\nu}=\partial_{\mu}e^{a}\,_{\nu} - \Gamma^{\lambda}\,_{\mu\nu}e^{a}\,_{\lambda}=0\,.
\end{equation}
The connection $\Gamma^{\lambda}\,_{\mu\nu}$ guarantees the parallelism and is not symmetric in the permutation of the last two indices. Isolating $\Gamma^{\lambda}\,_{\mu\nu}$ in Eq. (\ref{eq1}), we obtain
\begin{equation}\label{eq2}
\Gamma^{\lambda}\,_{\mu\nu}=e_{a}\,^{\lambda}\partial_{\mu}e^{a}\,_{\nu}\,.
\end{equation}
The above connection is the Weitzenb\"{o}ck connection and yields the non-null torsion tensor
\begin{equation}\label{eq3}
T^{\lambda}\,_{\mu\nu}=\Gamma^{\lambda}\,_{\mu\nu}-\Gamma^{\lambda}\,_{\nu\mu}\,.
\end{equation}
Using Eqs. (\ref{eq2}) and (\ref{eq3}), we arrive at
\begin{equation}\label{eq4}
T^{a}\,_{\mu\nu} = \partial_{\mu}e^{a}\,_{\nu} - \partial_{\nu}e^{a}\,_{\mu}\,.
\end{equation}

In TEGR, the frame is characterized from its acceleration $a^{(i)}$. Assuming the observer carries orthonormal tetrads $e^{a}\,_{\mu}$ along his world line $x^{\mu}(\tau)$ we may evaluate the acceleration of the frame as \cite{mashhoon2002length,mashhoon2003vacuum}
\begin{equation}\label{eq5}
a^{\mu} = u^{\alpha}\nabla_{\alpha}u^{\mu} = u^{\alpha}\nabla_{\alpha}e_{(0)}\,^{\mu}=\frac{d^{2}x^{\mu}}{d\tau^{2}}+\mathring{\Gamma}^{\mu}\,_{\alpha\beta}\frac{dx^{\alpha}}{d\tau}\frac{dx^{\beta}}{d\tau}\,,
\end{equation}
where $\mathring{\Gamma}^{\mu}\,_{\alpha\beta}$ are the Christoffel symbols.
It can be shown that \cite{maluf2013teleparallel}
\begin{equation}\label{eq6}
a^{(i)} = e^{(i)}\,_{\mu}a^{\mu} = \eta^{(i)b}\phi_{(0)b} = \frac{1}{2}T_{(0)(0)}\,^{(i)}\,,
\end{equation}
where $\phi_{a}\,^{b}=e^{b}\,_{\mu}u^{\lambda}\nabla_{\lambda}e_{a}\,^{\mu}$.
When the frame is in free fall, $u^{\mu}$ represents a geodesic and $a^{\mu}=0$.
Thus, we may identify $\phi_{ab} \leftrightarrow (\bf{a},\bf{\Omega})$, where $\bf{a}$ is the translational acceleration and $\bf{\Omega}$ the rotational frequency of the local spatial frame relative to a non-rotating transported frame \cite{mashhoon2002length,mashhoon2003vacuum}.
The establishment of the static nature of the observer is made at space infinity, for asymptotically flat spacetimes, by requiring that
\begin{equation}\label{eq7}
\lim_{r\rightarrow\infty}e^{a}\,_{\mu}=\delta^{a}_{\mu}\,,
\end{equation}
where $r$ represents the distance from the observer to the source of the gravitational field.

The Lagrangian formulation of the theory is constructed from the Lagrangian density \cite{maluf2013teleparallel}
\begin{equation}\label{eq8}
\mathcal{L}=-ke\Sigma^{abc}T_{abc}-\mathcal{L}_{M}\,,
\end{equation}
where $k=1/16\pi$, $\mathcal{L}_{M}$ is the Lagrangian density of the matter-radiation fields and 
\begin{equation}\label{eq9}
\Sigma^{abc}\equiv\frac{1}{4}\left(T^{abc}+T^{bac}-T^{cab}\right)+\frac{1}{2}\left(\eta^{ac}T^{b}-\eta^{ab}T^{c}\right)\,.
\end{equation}
The field equations derived from (\ref{eq8}) are \cite{maluf2013teleparallel}
\begin{equation}\label{eq10}
e_{a\lambda}e_{b\mu}\partial_{\nu}\left(e\Sigma^{b\lambda\nu}\right)-e\left(\Sigma^{b\nu}\,_{a}T_{b\nu\mu}-\frac{1}{4}e_{a\mu}T_{bcd}\Sigma^{bcd}\right)=\frac{1}{4k}eT_{a\mu}\,,
\end{equation}
where $T_{\mu\nu}=e^{a}\,_{\mu}T_{a\nu}$ is the mater-radiation energy-momentum tensor. 
The field equations (\ref{eq10}) are co-variant under coordinate transformations, local and global Lorentz transformations, and can be written as $R_{a\mu}(e)-\frac{1}{2}e_{a\mu}R(e) = \frac{1}{2k}T_{a\mu}$ \cite{maluf2013teleparallel}, where $R_{a\mu}$ and $R$ are, respectively, the projected Ricci tensor and Ricci scalar of the Riemann space constructed from the Weitzenb\"{o}ck torsion.

In the Hamiltonian formulation of TEGR, the gravitation energy-momentum is well established and can be formally obtained (see Ref. \cite{maluf2013teleparallel} for a complete discussion). However, the same result can be easily obtained directly from the Lagrangian formulation of the theory and suffices for the purpose of this paper. The field equations (\ref{eq10}) can be written as
\begin{equation}\label{eq11}
\partial_{\nu}\left(e\Sigma^{a\lambda\nu}\right)=\frac{1}{4k}ee^{a}\,_{\mu}\left(t^{\lambda\mu}+T^{\lambda\mu}\right)\,,
\end{equation}
where
\begin{equation}\label{eq12}
t^{\lambda\nu}=k\left(4\Sigma^{bc\lambda}T_{bc}\,^{\mu}-g^{\lambda\mu}\Sigma^{bcd}T_{bcd}\right)\,.
\end{equation}
Using the antisymmetry of $\Sigma^{a\lambda\nu}$, we obtain from Eq. (\ref{eq11}) 
\begin{equation}\label{eq13}
\partial_{\lambda}\left[ee^{a}\,_{\mu}\left(t^{\lambda\mu}+T^{\lambda\mu}\right)\right]=0\,.
\end{equation}
Integrating (\ref{eq13}), we obtain the continuity equation
\begin{equation}\label{eq14}
\frac{d}{dx^{0}}\int_{V}{d^{3}x\left[ee^{a}\,_{\mu}\left(t^{0\mu}+T^{0\mu}\right)\right]}=-\oint_{S}{dS_{i}\left[ee^{a}\,_{\mu}\left(t^{i\mu}+T^{i\mu}\right)\right]}\,.
\end{equation}
Thus
\begin{equation}\label{eq15}
P^{a}\equiv\int_{V}{d^{3}xee^{a}\,_{\mu}\left(t^{0\mu}+T^{0\mu}\right)}\,,
\end{equation}
represents a conserved quantity. Identifying $t^{\mu\nu}$ as the gravitational energy-momentum tensor in Eq. (\ref{eq11}), we obtain
\begin{equation}\label{eq16}
P^{a}=4k\int_{V}{d^{3}x \, \partial_{i}(e\Sigma^{a0i}})=4k\oint_{S}dS_{i}(e\Sigma^{a0i})\,,
\end{equation}
where $S$ is the closed surface encompassing V. The quantity in the above equation represent the total energy-momentum four-vector contained in the volume V. Thus from Eq. (\ref{eq15}), we identify 
\begin{equation}\label{eq17}
P_{g}^{a}=\int_{V}{d^{3}x \, e(e^{a}\,_{\mu}t^{0\mu}})
\end{equation}
as the gravitational energy-momentum four-vector contained in the volume $V$, and
\begin{equation}\label{eq18}
P_{m}^{a}=\int_{V}{d^{3}x \, (ee^{a}\,_{\mu}T^{0\mu}})
\end{equation}
as the energy-momentum four-vector of the source contained in the volume $V$.
From the spatial components of Eq. (\ref{eq14}), we define the gravitation energy-momentum flux as
\begin{equation}\label{eq19}
\Phi^{a}_{g}= \oint_{S}dS_{i} \, (ee^{a}\,_{\mu}t^{i\mu})\,.
\end{equation}
The four-vector $P^{a}$ is invariant under coordinate and global Lorentz transformations and transforms as a four-vector under $SO(3,1)$ symmetry. The zero component $P^{(0)}$ represents the energy as in Special Relativity and is frame dependent. As mentioned before, Eq. (\ref{eq16}) is obtained from the vacuum equations of the Hamiltonian formulation of the TEGR \cite{maluf2013teleparallel}; and the same expression is obtained through the Noether's charges \cite{emtsova2021conserved}.

In sections \ref{static} and \ref{freefall} we evaluate the quantities (\ref{eq16}-\ref{eq19}) for two distinct observers.

\section{The Alcubierre spacetime}\label{review}

Alcubierre spacetime is a non-trivial geometry formed by a shell, where the spacetime inside the shell is conformally flat and the exterior is Minkowski spacetime. The shell is denoted as a ``bubble''. The spacetime is asymptotically flat \cite{mcmonigal2012alcubierre}, therefore regularly represented by the coordinates $x^{\mu}=(t,x,y,z)$ at space infinity. The bubble propagates with a global velocity $v_{s}$, which can be greater than unity in geometrized units, in the positive $z$ direction. The coordinates $x^{\mu}$ are adapted to a static observer that measures the velocity $v_{s}$ and the position $r_{s}$ for the bubble, where $v_{s}=dr_{s}/dt$. Thus, the quantity $r_{s}$ localize the bubble's center in relation to a static observer. In the $x^{\mu}$ coordinate system the line element may be written as \cite{alcubierre2017warp}
\begin{equation}\label{eq20}
ds^{2}=\big(f_{s}^{2}v_{s}^{2}-1\big) \, dt^{2}+dx^{2}+dy^{2}+dz^2-2f_{s}v_{s}dtdz\,,
\end{equation}
where $r_{s}=\sqrt{(z-v_{s}t)^{2}+x^{2}+y^{2}}$ and $f_{s}=f_{s}(r_{s})$. The function $f_{s}$ is the non-trivial geometric term and it may be, arbitrarily, chosen as long as some conditions are satisfied, i.e., $f_{s}=constant$ inside the bubble and $f_{s}=0$ outside of the bubble. The spacetime are of Type $\mathcal{I}$ in Petrov classification and there is no curvature singularities \cite{mattingly2021curvature}. For the sake of simplicity, we shall introduce polar coordinates into the two-dimensional space transverse to the propagation direction. Thus, the line element (\ref{eq20}) reads
\begin{equation}\label{eq21}
ds^{2}=\big(f_{s}^{2}v_{s}^{2}-1\big) \, dt^{2}+d\rho^{2}+\rho^{2}d\phi^{2}+dz^2-2f_{s}v_{s}dtdz\,,
\end{equation}
where $r_{s}=\sqrt{(z-v_{s}t)^{2}+\rho^{2}}$, $x=\rho\cos{\phi}$ and $y=\rho\sin{\phi}$. In the $(t,\rho,\phi,z)$ coordinates, the non-zero components $G^{\mu\nu}$ of the Einstein tensor are
\begin{align}
G^{00}&=-\frac{1}{4}\rho^{2} v_{s}^{2} \left(\frac{f^{'}_{s}}{r_{s}}\right)^{2}\,,\label{eq22}\\
G^{01}&=-\frac{1}{2}\frac{\rho v_{s}(tv_{s}-z)}{r_{s}}\frac{d}{dr_{s}}\left(\frac{f^{'}_{s}}{r_{s}}\right)\,,\label{eq23}\\
G^{03}&=-\frac{1}{4}v_{s}\frac{\rho ^2 v_{s}^2 r_{s} f_{s} f^{'2}_{s}+2 \left(\rho ^2+2 t^2 v_{s}^2-4t v_{s} z+2 z^2\right) f^{'}_{s}+2 \rho ^2 r_{s} f^{''}_{s}}{r_{s}^{3}}\,,\label{eq24}\\
G^{11}&=\frac{1}{4}\frac{v_{s}^{2}}{r_{s}^{2}}\bigg[\frac{4\rho^{2}}{r_{s}}(1-f_{s}-)f^{'}_{s} + 2\rho^{2}f^{'2}_{s}+ \big[4(t v_{s}-z)^{2} -\rho^{2} \big]+4(t v_{s}-z)^{2} (f_{s} - 1)f^{''}_{s} \bigg]\,,\label{eq25}\\
G^{13}&=-\frac{1}{2}\frac{\rho(t v_{s}-z)}{r^{2}_{s}}\bigg[(1-2f_{s})f^{'}_{s}+ 2r_{s}f^{'2}_{s}+r_{s}(2f_{s}-1)f^{''}_{s}\bigg]\,,\label{eq26}\\
G^{22}&=-\frac{1}{4}\frac{v_{s}}{r_{s}^{3}}\bigg[ 4\rho^{2}(f_{s}-1)f^{'}_{s}+r_{s}\big[4(-t v_{s}+z)^{2} +\rho^{2} \big]f^{'2}_{s} +4(-t v_{s} + z)^{2} r_{s}(f_{s}-1)f^{''}_{s} \bigg]\,,\label{eq27}\\
G^{33}&=-\frac{1}{4}\frac{v_{s}^{2}}{r_{s}^{3}}\bigg[ 3\rho^{2} r_{s} f^{'2}_{s} + v^{2}_{s}\rho^{2}r_{s}f_{s}^{2}f_{s}^{'2} + 4f_{s}\big[ (2(-t v_{s}+z)^{2} +\rho^{2}) f^{'}_{s} +\rho^{2}r_{s}f^{''}_{s} \big] \bigg]\,,\label{eq28}
\end{align}
where the prime indicates a derivation with respect to $r_{s}$.

A test particle inside the bubble experiences a geodesic motion, i.e., it is dragged with the bubble to any value of $v_{s}$ \cite{alcubierre2017warp}. The four-velocity of this particle is \cite{alcubierre2017warp}
\begin{equation}\label{eq29}
u^{\mu}=(1,0,0,v_{s}f_{s})\,.
\end{equation}
The time dilatation effect on the particle can be obtained simply by considering this particle along the propagation axis, i.e., $\rho=0$, in the line element (\ref{eq21}) and $f_{s}=1$. Therefore, in the center of the bubble $dz=v_{s}dt$, thus we have $d\tau=dt$. Note that realistically minor dilatation effects may occur, but can be controlled by choosing a bubble large or thin enough.

The expansion of timelike congruence can be defined as $\Theta=\partial_{\mu}u^{\mu}$ \cite{poisson2004relativist}, where $u^{\mu}$ is the tangent field to the congruence. Hence, using (\ref{eq29}), we obtain \cite{alcubierre2017warp}
\begin{equation}\label{eq30}
\Theta=v_{s}(\partial_{z}r_{s})f'_{s}=v_{s}\frac{z-z_{0}}{r_{s}}f^{'}_{s}\,,
\end{equation}
where $z_{0}=v_{s}t$ is the center of the bubble. We can see that in the front of the bubble ($z>z_{0}$) we have $\Theta>0$ and in the rear ($z<z_{0}$) we have $\Theta<0$. Hence, the Alcubierre spacetime may be interpreted as a ``expansion'' of the spacetime followed by a ``contraction''. The expansion was latter shown by Nat{\'a}rio not to be a necessary property \cite{natario2002warp} in the construction of his ``Nat{\'a}rio drives'' as now called.

Some properties required for the source of this gravitational field can be perceived through Einstein equations. The source energy-momentum component $T^{00}$ reads
\begin{equation}\label{eq31}
T^{00}=2kG^{00}=-\frac{k}{2}\rho^{2} v_{s}^{2} \left(\frac{f^{'}_{s}}{r_{s}}\right)^{2}\leq 0\,.
\end{equation}
We can see that this component implies a negative energy density, in spacetime, for the source. 
In TEGR, the energy is the $(0)$ component of a four-vector and depends on the frame. Thus, in order to measure a physical property, we must establish a frame, i.e., an observer. In section \ref{conclusions}, we compare the results obtained in section \ref{static} and \ref{freefall} and address the ``negative energy problem''. It can be seen that when $v_{s}=v_s(t)$ the component $T^{00}$ varies with time. We address this in section \ref{vt}.

The function $f_{s}(r_{s})$ may assume different forms, but in general it has the dependency on two parameters related to the radius of the bubble and its thickness. The original function considered by Alcubierre is \cite{alcubierre1994warp}
\begin{equation}\label{eq32}
f_{s}(r_{s})=\frac{\tanh{[\sigma(r_{s}+R)]-\tanh{[\sigma(r_{s}-R)]}}}{2\tanh{\sigma R}}\,,
\end{equation}
where the parameter $R$ is related to the radius of the bubble and $\sigma$ to the inverse of its thickness. Thus, we have three regions I, II and III. I and III are flat and separated by the non-flat region II.

In this paper, we shall consider the function given by (\ref{eq32}) and, for comparison, another form given by
\begin{equation}\label{eq33}
f_{s}(r_{s})=e^{-\sigma \, r_{s}^{2}}\,,
\end{equation}
where $\sigma$ is a term that regulates the smoothness of the transition between the wall and the flat region. 
For the function (\ref{eq33}), the bubble wall consumes almost all of its interior, leaving the spacetime flat only in the vicinity of the center of the bubble.

\section{Static observer}\label{static}

In order to evaluate the energy of the spacetime, we need to establish a tetrad field that is associated with the spacetime metric and adapted to an observer. A frame of reference can be characterized by a congruence of stationary observers (for instance, at spatial infinity).  Each observer carries along its worldline a set of tetrads that are orthonormal fields.  Then at the limit of $r\rightarrow\infty$ the tetrad field takes the form $e^a\,_\mu\rightarrow\delta^a_\mu$ for stationary observers.  The component $e_{(0)}\,^\mu$ describes a congruence of time-like vectors and is associated with the velocity field of the observer $u^\mu$.  Thus a field of observers is described by its velocity field and by the acceleration of the frame as a whole.  Such acceleration is given by the absolute derivative of the tetrads with respect to an affine parameter such as the proper time. That is
\begin{equation}
\frac{De_a\,^\mu}{d\tau}=\phi_a\,^b\,e_b\,^\mu\,,
\end{equation}
where $\phi_{ab}$ is the acceleration tensor. We adopt the interpretation presented in the reference [18], in which $\phi_{(0)(i)}$ is associated to the translational acceleration of the frame, while $\phi_{(i)(j)}$ is the frequency of rotation with respect to a non-rotating Fermi- Walker transported frame. As it was stated after eq. (6).
In this section we consider a static observer, i.e., $u^{i}=e_{(0)}\,^{i}=0$. A static observer will only measure the physical properties of the spacetime and not quantities related to his motion. A set of tetrads that satisfies the requirements is 
\begin{equation}\label{eq34}
e_{a\mu}=\left(
\begin{array}{cccc}
 -A & 0 & 0 & -B \\
 0 & \cos{\phi} & -\rho \sin{\phi} & 0\\
 0 & \sin{\phi} & \rho \cos{\phi} & 0 \\
 0 & 0 & 0 & 1/A \\
\end{array}
\right)\,,
\end{equation}
where $A=\sqrt{1-f_{s}^{2}v_{s}^{2}}$ and $B=f_{s}v_{s}/A$. 
The inverse tetrads read
\begin{equation}\label{eq35}
e^{a\mu}=\left(
\begin{array}{cccc}
 -1/A & 0 & 0 & 0 \\
 0 & \cos{\phi} & -\sin{\phi}/\rho & 0\\
 0 & \sin{\phi} & \cos{\phi}/\rho & 0 \\
 -B & 0 & 0 & A \\
\end{array}
\right)\,.
\end{equation}
The determinant is $e=\rho$. It is easy to check that $e_{(0)}\,^{\mu}=(1/A,0,0,0)$.
At space infinity, we have $\displaystyle{\lim_{r\rightarrow\infty}}e_{a}\,^{\mu}=\delta_{a}^{\mu}$, where $r\equiv\sqrt{\rho^{2}+z^{2}}$, i.e., we have Minkowski spacetime. The the tetrads (\ref{eq34}) are adapted to a static observer. The metric tensor can be easily recovered remembering that $g_{\mu\nu}=e_{a\mu}e^{a}\,_{\nu}$.
The non-vanishing components of the torsion tensor $T^{\lambda\mu\nu}$ can be calculated using (\ref{eq4}). They read
\begin{align}
T^{001}&=\frac{\rho v_{s}^{2}}{r_{s}}f_{s}f^{'}_{s}\,,\label{eq36}\\
T^{003}&=-\frac{v_{s}^{2}(tv_{s}-z)}{r_{s}A^{2}}(1-f_{s}+v_{s}^{2}f_{s}^{3})f^{'}_{s}\,,\label{eq37}\\
T^{013}&=-\frac{\rho v_{s}}{r_{s}A^{2}}f^{'}_{s}\,,\label{eq38}\\
T^{301}&=-\frac{\rho v_{s}^{3}}{r_{s}A^{2}}f^{2}_{s}f^{'}_{s}\,,\label{eq39}\\
T^{303}&=\frac{v_{s}^{3}(tv_{s}-z)}{r_{s}A^{2}}f_{s}^{2}f^{'}_{s}\,,\label{eq40}
\end{align}
where, again, the prime indicates a derivation with respect to $r_s$.

The only non-vanishing component $e\Sigma^{(0)01}$ constructed from (\ref{eq9}) and (\ref{eq36}-\ref{eq40}) is
\begin{equation}\label{eq41}
e\Sigma^{(0)01}=-\frac{\rho^{2}v_{s}^{2}f_{s}f'_{s}}{4Ar_{s}}\,.
\end{equation}
Therefore, from (\ref{eq16}), choosing a very long cylinder as the surface of integration,  we obtain the energy
\begin{align}\label{eq42}
P^{(0)}&=-k\int_{S}dS_{1} \, \Bigg(\frac{\rho_{0}^{2}v_{s}^{2}f_{s}f'_{s}}{r_{s}A}\Bigg)\nonumber\\
&=-\frac{\rho_{0}^{2}}{8}\int_{-L}^{L}dz \, \Bigg(\frac{v_{s}^{2}f_{s}f'_{s}}{r_{s}A}\Bigg)\,.
\end{align}
Eq. (\ref{eq42}) represents the total energy contained inside a spatial region of $\rho=\rho_{0}=constant$, i.e., a cylinder of radius $\rho_{0}$ and length $2L$ in Euclidean space. In order to evaluate only the gravitational energy, we need the components $t^{\mu\nu}$ given by (\ref{eq12}). They read
\begin{align}
t^{00}&=-k\frac{\rho ^2 v_{s}^2 }{2 r_{s}^{2}A^{2}}\left(1+v_{s}^{2}f_{s}^{2}\right)f_{s}^{'2}\,,\label{eq43}\\
t^{03}&=-k\frac{\rho^{2}v_{s}^{3}}{2r_{s}^{2}A^{2}} f_{s}  \left(1+v_{s}^2 f^{2}_{s}\right)f^{'2}_{s}\label{eq44}\,.
\end{align}
Then, from (\ref{eq17}), we have
\begin{align}
P_{g}^{(0)}&=-\frac{k}{2}\int_{V}d^{3}x \, \Bigg(\frac{\rho^{3} v_{s}^2  \left(1+v_{s}^2 f_{s}^2 \right)f^{'2}_{s}}{r_{s}^{2} A^{3}}\Bigg)\nonumber\\
&=-\frac{k}{2}\int_{0}^{2\pi}d\phi\int_{-L}^{L}dz\int_{0}^{\rho_{0}}d\rho \, \Bigg(\frac{\rho^{3} v_{s}^2  \left(1+v_{s}^2 f_{s}^2 \right)f^{'2}_{s}}{r_{s}^{2} A^{3}}\Bigg)\,.\label{eq45}
\end{align}
The energy of the source, measured by the static observer, can be evaluated using (\ref{eq18}) with the aid of Einstein equations, (\ref{eq22}-\ref{eq28}) and (\ref{eq34}). Thus,
\begin{align}
P_{m}^{(0)}&=-\int_{V}d^{3}x \, e\left(e_{(0)0}T^{00}+e_{(0)3}T^{03}\right)\nonumber\\
&=-2k\int_{V}d^{3}x \, e\left(e_{(0)0}G^{00}+e_{(0)3}G^{03}\right)\nonumber\\
&=-\frac{k}{2}\int_{0}^{2\pi}d\phi\int_{-L}^{L}dz\int_{0}^{\rho_{0}}d\rho \, \Bigg(\rho v_{s}^{2}\frac{\rho^{2}r_{s}f^{'2}_{s}+2f_{s}[\rho^{2}r_{s}f^{''}_{s}+(2r_{s}^{2}-\rho^{2})f_{s}^{'}]}{r_{s}^{3}A}\Bigg)\label{eq46}\,.
\end{align}
The consistency of the calculations can be verified analytically by adding Eqs. (\ref{eq45}) and (\ref{eq46}). By doing it, we can easily obtain (\ref{eq42}), so the Eq. (\ref{eq11}) is satisfied. The relation $\partial_{\rho}f_{s}=(\rho/r_{s})f^{'}_{s}$ may be useful in achieving such an equality.

\begin{figure}
\centering
\begin{minipage}{.47\textwidth}
	\centering
		\includegraphics[width=1\textwidth]{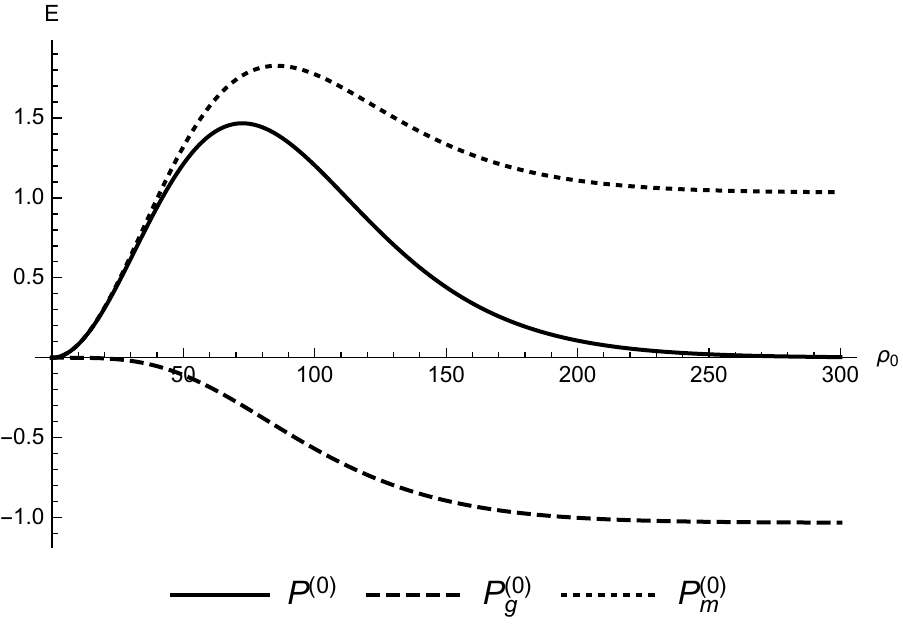}
	\caption{Total (\ref{eq42}), gravitational (\ref{eq45}) and source (\ref{eq46}) energies $E$ measured by a static observer for different radii $\rho_{0}$. The integrals were calculated over an infinity cylinder of radius $\rho_{0}$. The continuous curve represents the total energy, the dashed the gravitational and the dotted the source. The quantity $\rho_{0}$ is a parameter, not a variable. The remaining parameters are $R=1$, $\sigma=0.01$, $t=0$, $v_{s}=0.5$ and $f_{s}$ given by (\ref{eq32}).}
	\label{fig1}
\end{minipage}
\qquad
\begin{minipage}{.47\textwidth}
		\includegraphics[width=1\textwidth]{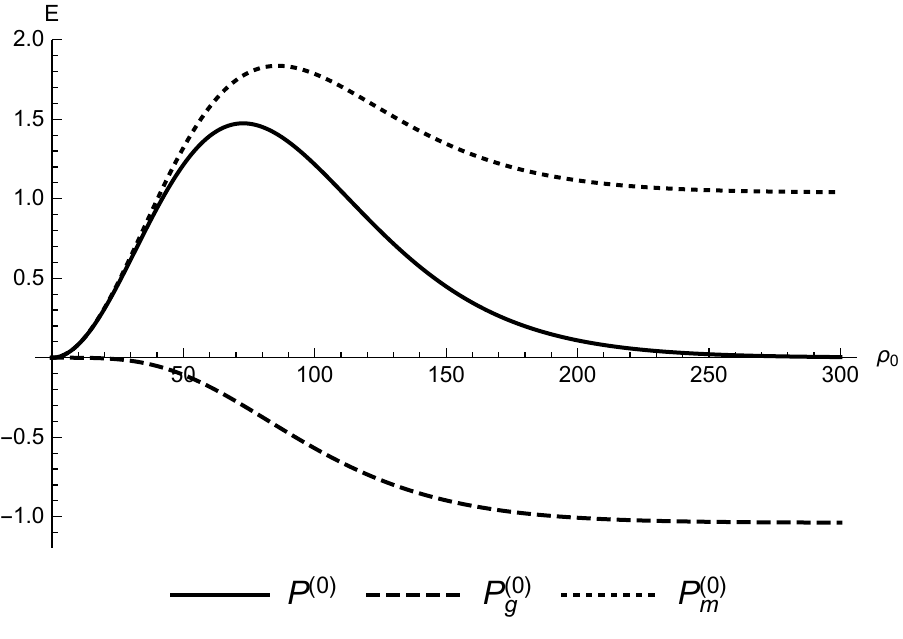}
	\caption{Total (\ref{eq42}), gravitational (\ref{eq45}) and source (\ref{eq46}) energies $E$ measured by a static observer for different radii $\rho_{0}$. The integrals were calculated over an infinity cylinder of radius $\rho_{0}$. The continuous curve represents the total energy, the dashed the gravitational and the dotted the source. The quantity $\rho_{0}$ is a parameter, not a variable. The remaining parameters are $R=50$, $\sigma=0.01$, $t=0$, $v_{s}=0.5$ and $f_{s}$ given by (\ref{eq32}).}
	\label{fig2}
\end{minipage}
\end{figure}

\begin{figure}
\centering
\begin{minipage}{.47\textwidth}
	\centering
		\includegraphics[width=1\textwidth]{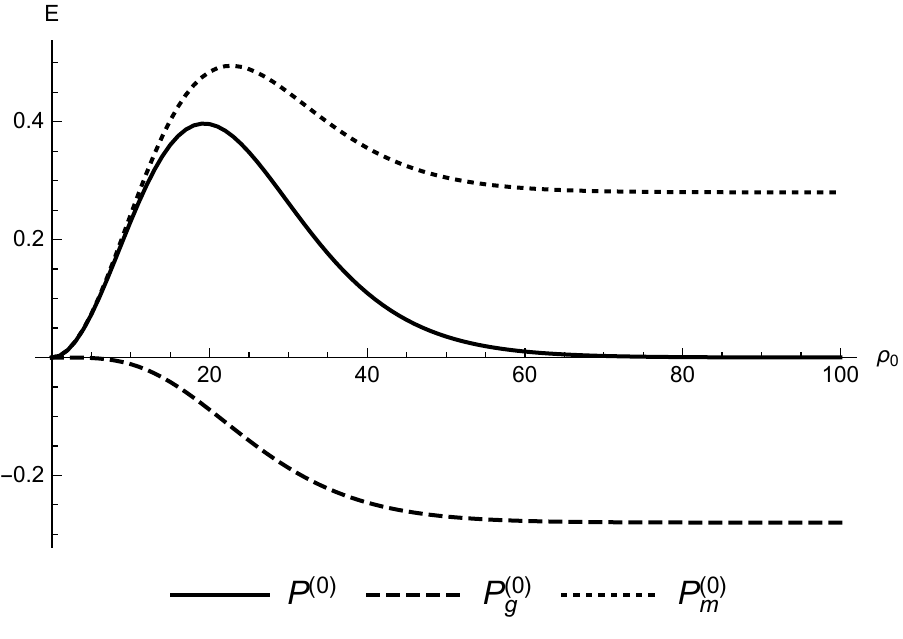}
	\caption{Total (\ref{eq42}), gravitational (\ref{eq45}) and source (\ref{eq46}) energies $E$ measured by a static observer for different radii $\rho_{0}$. The integrals were calculated over an infinity cylinder of radius $\rho_{0}$. The continuous curve represents the total energy, the dashed the gravitational and the dotted the source. The quantity $\rho_{0}$ is a parameter, not a variable. The remaining parameters are $R=1$, $\sigma=0.04$, $t=0$, $v_{s}=0.5$ and $f_{s}$ given by (\ref{eq32}).}
	\label{fig3}
\end{minipage}
\qquad
\begin{minipage}{.47\textwidth}
	\centering
		\includegraphics[width=1\textwidth]{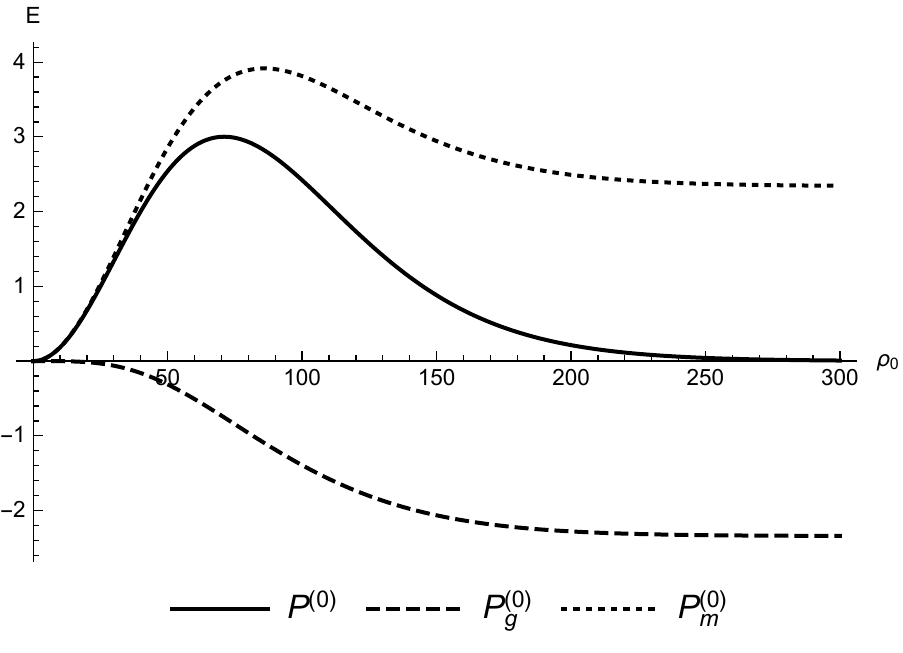}
	\caption{Total (\ref{eq42}), gravitational (\ref{eq45}) and source (\ref{eq46}) energies $E$ measured by a static observer for different radii $\rho_{0}$. The integrals were calculated over an infinity cylinder of radius $\rho_{0}$. The continuous curve represents the total energy, the dashed the gravitational and the dotted the source. The quantity $\rho_{0}$ is a parameter, not a variable. The remaining parameters are $R=1$, $\sigma=0.01$, $t=0$, $v_{s}=0.7$ and $f_{s}$ given by (\ref{eq32}).}
	\label{fig4}
\end{minipage}
\end{figure}

\begin{figure}
\centering
	\centering
		\includegraphics[width=.6\textwidth]{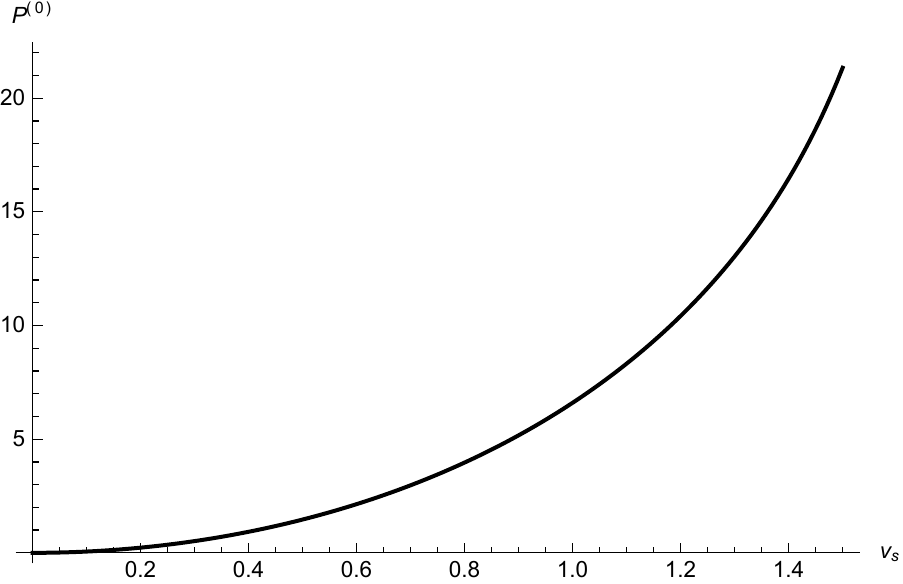}
	\caption{Total energy (\ref{eq42}) measured by a static observer for different $v_{s}$. The integral was calculated over an infinity cylinder of radius $\rho_{0}=75$. The quantity $v_{s}$ is a parameter, not a variable. The parameters are $R=1$, $t=0$, $\sigma=0.01$ and $f_{s}$ given by (\ref{eq32}).}
	\label{fig5}
\end{figure}

The integrals (\ref{eq42},\ref{eq45},\ref{eq46}) can be evaluated numerically and the results seen in Figure \ref{fig1} for distinct radius $\rho_{0}$ using the original Alcubierre function (\ref{eq32}).
We can see that the gravitational and source energies converge to constant values at space infinity, but the total energy is zero at space infinity. This is an interesting behavior never before observed in TEGR. The total energy of the Alcubierre spacetime is zero at infinity, but not around the ship. We can see a peak and then a progressive decrease.
Also interesting, the energy of the source is positive, while the gravitational is negative. Negative gravitational energies are not a new feature in TEGR, already obtained by integrating the Schwarzschild spacetime from the event horizon to infinity; and for the spacetime of plane gravitational waves \cite{maluf2008energy}. 
The effect of the radius $R$ on (\ref{eq32}) can be perceived by comparing Figures \ref{fig1} and \ref{fig2}. When increasing the radius, the energy also increases, but the qualitative behavior is not altered.
The energy is more sensitive to the thickness parameter $\sigma$ of the bubble than to its radius $R$, as we can see a significant reduction in energy for a smoother $\sigma$ in Figure \ref{fig3}.
Thus, a larger bubble with a thin border appears to be more ``energy efficient''. The energy distribution seems to be more localized for a thin bubble, as the total energy tends to zero in a smaller radius.
The energy has a strong dependence on the velocity $v_{s}$ as can be perceived from Figure \ref{fig4}. With an increase of only approximately $40\%$ in the velocity, the energy approximately doubles.
This increase in the energy with the velocity can be better seen in Figure \ref{fig5}, where a fixed integration radius $\rho_{0}$ is chosen, and the total energy is evaluated for different values of $v_{s}$. In Figure \ref{fig5}, $\rho_{0}$ was chosen to be the radius around the peak of the energy. The position of the peak does not depend on $v_{s}$.

\begin{figure}
	\centering
		\includegraphics[width=.6\textwidth]{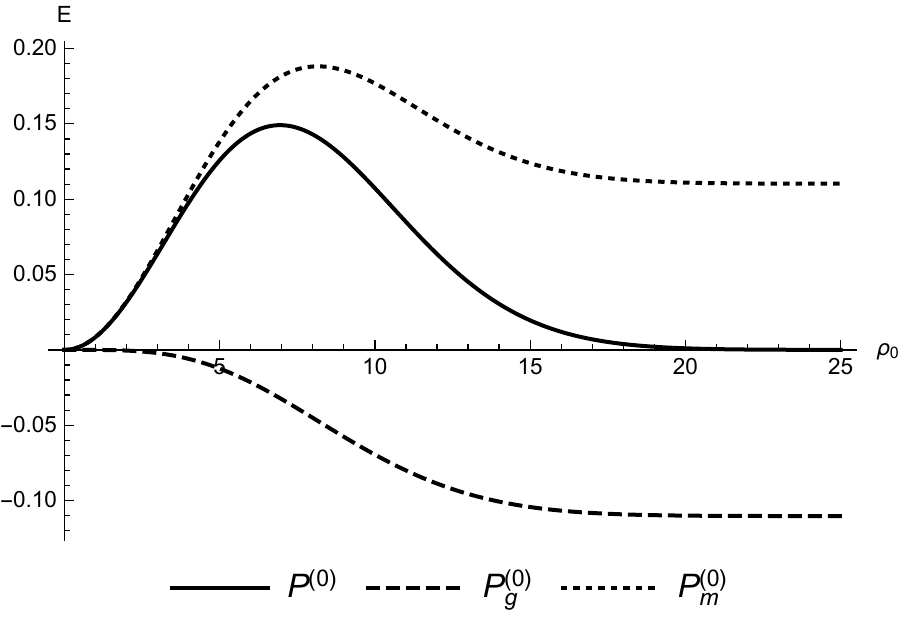}
	\caption{Total (\ref{eq42}), gravitational (\ref{eq45}) and source (\ref{eq46}) energies $E$ measured by a static observer for different radii $\rho_{0}$. The integrals were calculated over an infinity cylinder of radius $\rho_{0}$. The continuous curve represents the total energy, the dashed the gravitational and the dotted the source. The quantity $\rho_{0}$ is a parameter, not a variable. The remaining parameters are $\sigma=0.1$, $t=0$, $v_{s}=0.5$ and $f_{s}$ given by (\ref{eq33}).}
	\label{fig6}
\end{figure}
The energy also depends on the form of the function $f_{s}$, as we can see in Figure \ref{fig6} that the smoother function (\ref{eq33}) greatly reduces energy requirements.

\subsection{Gravitational radiation measured by a static observer}\label{static_radiation}

\begin{figure}
\centering
\begin{minipage}{.47\textwidth}
	\centering
		\includegraphics[width=1\textwidth]{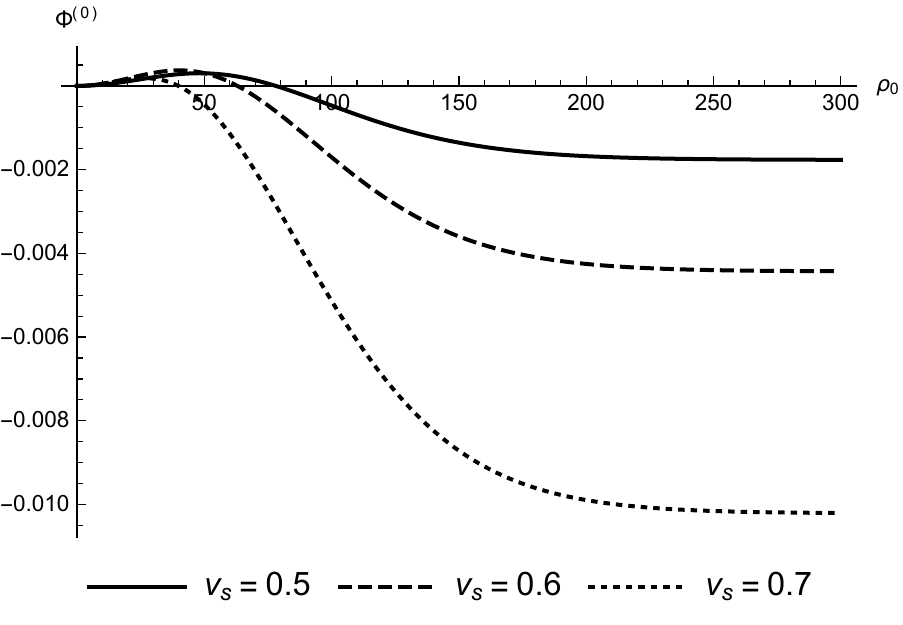}
	\caption{Gravitational flux (\ref{eq49}) measured by a static observer for different radii $\rho_{0}$. The integrals were calculated over a cylinder of radius $\rho_{0}$ and length $2L$. The quantity $\rho_{0}$ is a parameter, not a variable. The remaining parameters are $R=1$, $\sigma=0.01$, $t=100$, $L=100$, and $f_{s}$ given by (\ref{eq32}).}
	\label{fig7}
\end{minipage}
\qquad
\begin{minipage}{.47\textwidth}
	\centering
		\includegraphics[width=1\textwidth]{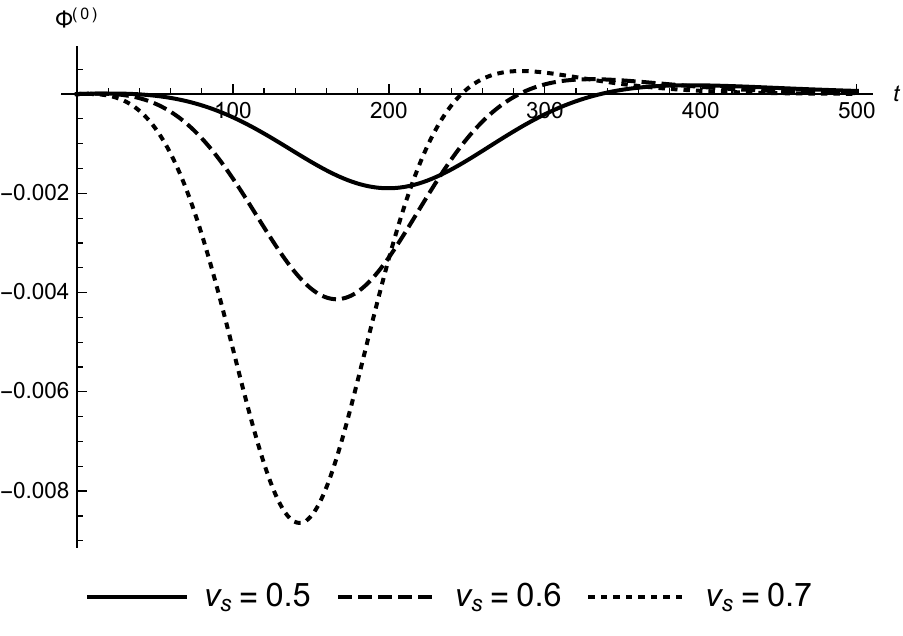}
	\caption{Gravitational flux (\ref{eq49}) measured by a static observer as a function of $t$. The integrals were calculated over a cylinder of radius $\rho_{0}$ and length $2L$. The parameters are $R=1$, $\sigma=0.01$, $L=100$, $\rho_{0}=100$, and $f_{s}$ given by (\ref{eq32}).\\}
	\label{fig8}
\end{minipage}
\end{figure}

In this subsection, we evaluate the gravitational flux measured by a static observer, e.g., an observer on Earth that measures the gravitational flux as the bubble moves away. As we are considering a ship moving in the positive $z$ direction, the static observer is always behind the ship for $t>0$. Defining the quantity $\phi_{g}^{aj}=e^{a}\,_{\mu}t^{j\mu}$ we may write Eq. (\ref{eq19}) as $\Phi_{g}^{a}=\oint_{S}dS_{j}(e\phi^{aj}_{g})$. The evaluation of the components $\phi_{g}^{(0)i}$ is straightforward from the torsion tensor components (\ref{eq36}-\ref{eq40}). We obtain
\begin{align}
\phi_{g}^{(0)1}&=k\frac{\rho^{2} v_{s}^{3}(tv_{s}-z)}{r_{s}^{2}A^{3}}\big[1-f_{s}(1+v_{s}^{2}f_{s}^{2}) \big]f_{s}^{'2}\,,\label{eq47}\\
\phi_{g}^{(0)3}&=-k\frac{\rho^{3} v_{s}^{3}}{r_{s}^{2}A^{3}}f_{s}\big(1+v_{s}^{2}f_{s}^{2}\big)f^{'2}_{s}\,,\label{eq48}
\end{align}
as the only non-zero components. The component $\Phi_{g}^{(0)}$ represents the gravitational energy flowing through a finite closed surface. Observing the components (\ref{eq47},\ref{eq48}) we may perceive that we have gravitational energy flux normal to the surfaces $dS_{1}$ and $dS_{3}$, i.e., through a ``cylinder'' with sides surface, base and top. 
The total flux reads
\begin{align}
\Phi_{g}^{(0)}&=\int dS_{1}\phi_{g}^{(0)1}+\int dS_{3}\phi_{g}^{(0)3}\nonumber\\
&=-\frac{1}{8}\Bigg(\int_{0}^{\rho_{0}}d\rho \, \frac{\rho^{3} v_{s}^{3}}{r_{s}^{2}A^{3}}f_{s}\big(1+v_{s}^{2}f_{s}^{2}\big)f^{'2}_{s}\Bigg|_{z=L}
+\int_{0}^{\rho_{0}}d\rho \, \frac{\rho^{3} v_{s}^{3}}{r_{s}^{2}A^{3}}f_{s}\big(1+v_{s}^{2}f_{s}^{2}\big)f^{'2}_{s}\Bigg|_{z=-L}\nonumber\\
&- \int_{-L}^{L}dz \, \frac{\rho_{0}^{2} v_{s}^{3}(tv_{s}-z)}{r_{s}^{2}A^{3}}\big[1-f_{s}(1+v_{s}^{2}f^{2}) \big]f_{s}^{'2}
\Bigg)\,.\label{eq49}
\end{align}
Expression (\ref{eq49}) can be numerically integrated and the results are presented in Figures \ref{fig7} for different values of the velocity $v_{s}$.
A static observer measures a gravitational energy flux that increases (its absolute value) with the radius of the cylinder until it reaches a constant value. In Figure \ref{fig8} the gravitational energy flux is plotted over a fixed cylindrical surface as a function of time. 
The gravitational energy flux decays to zero as time increases, which is expected as the bubble has localized gravitational energy around itself and moves away from the observer. The expressions for the flux are not regular for $v_{s}>1$ and we believe that this feature is due to causality problems, since a static observer cannot measure radiation from a supraluminal source. In subsection \ref{geodesic_radiation}, we demonstrate that an observer moving with the bubble does not have this problem. The flux has a peak in time. The higher the velocity, the sooner the peak occurs. Since we are considering a surface fixed to the static observer, an observer moving with the bubble would see the rear surface moving away and the front moving closer. The peak occurs the moment that the bubble crosses the front surface. Thus, this increase in flux is caused by the bubble itself ``flowing'' through the surface.

\section{Free-falling observer}\label{freefall}

In section \ref{static} we obtained the gravitational energy and the total energy for the Alcubierre spacetime measured by a static observer.
In this section, we discuss the gravitational energy and source energy of Alcubierre spacetime measured by a free-falling observer, i.e., an observer that is in geodesic motion. Such an observer moves with the bubble, so the results obtained in this section can be interpreted as the results that an observer ``inside'' the bubble measures. In order to accomplish this task, one can change the coordinates by removing its origin from Earth and resetting it with the center of the bubble as its origin. In this new coordinate system, the surface equivalent to that used in (\ref{eq42}) would be considered to travel backward relative to the origin. However, the tetrad approach allows the analysis to take place in the same coordinate system, as we can transform the frames independently of the coordinates.
Applying a Lorentz transformation
\begin{equation}\label{eq50}
\Lambda^{a}\,_{b}=\left(\begin{array}{cccc}
 \gamma & 0 & 0 & \gamma v \\
 0 & 1 & 0 & 0 \\
  0 & 0 & 1 & 0 \\
	\gamma v & 0 & 0 & \gamma \\
\end{array}
\right)\,,
\end{equation}
in the positive direction of the $z$ axis in the tetrads (\ref{eq35}) we obtain a new set of tetrads $e^{'a}\,_{\mu}=\Lambda^{a}\,_{b}e^{b}\,_{\mu}$ that is adapted to an observer with velocity $v$ relative to the static observer and the origin of the coordinate system remains fixed on Earth, where $\gamma=(1-v^{2})^{1/2}$.
In order to determine the value of $v$ that leads to a free-falling observer, we evaluate the components of the accelerating tensor (\ref{eq6}) for the tetrads $e^{'a}\,_{\mu}$. They read
\begin{align}
\phi^{'(0)(1)}&=-\frac{\gamma^{2}v_{s}\rho}{r_{s}A^{2}}\big[v-v_{s}(1+v^{2})f_{s} +v v_{s}^{2}f_{s}^{2}\big]f^{'}_{s}\cos{\phi}\,,\label{eq51}\\
\phi^{'(0)(2)}&=-\frac{\gamma^{2}v_{s}\rho}{r_{s}A^{2}}\big[v-v_{s}(1+v^{2})f_{s} +v v_{s}^{2}f_{s}^{2}\big]f^{'}_{s}\sin{\phi}\,,\label{eq52}\\
\phi^{'(0)(3)}&=-\frac{\gamma^{3} (tv_{s}-z)}{r_{s}A^{3}}\Bigg\{v^2 (1-v^2) \big[1+v_{s}^2 f_{s}^3-f_{s} (1+v_{s}v)\big] f^{'}_{s}-A^{2}\big[v_{s}-(1+v_{s}^2 f-v_{s}^2 f^2) v\big] v' \Bigg\}\label{eq53}\,,
\end{align}
where $v'\equiv dv/dr_{s}$.
The above components are the components of the inertial acceleration necessary for an observer to remain in its dynamical state.
For a static observer ($v=0$), these are the components of the acceleration necessary for the observer to remain static and must vanish at space infinity, i.e., there is no inertial acceleration at infinity. Thus, we may obtain the value $v$ necessary for the observer to be in free fall requiring that all the components (\ref{eq51}-\ref{eq53}) vanish. It is straightforward to see that $\phi_{(0)1}=0$ and $\phi_{(0)2}=0$ lead to $v=v_{s}f_{s}$, and this also satisfy $\phi_{(0)3}=0$. Thus, the tetrads
\begin{equation}\label{eq54}
e^{'}_{a\mu}=\lambda_{a}\,^{b}e_{b\mu}=\left(
\begin{array}{cccc}
 -1 & 0 & 0 & 0 \\
 0 & \cos{\phi} & -\rho\sin{\phi} & 0\\
 0 & \sin{\phi} & \rho\cos{\phi} & 0 \\
 -v_{s}f_{s} & 0 & 0 & 1 \\
\end{array}
\right)\
\end{equation}
are adapted to a free-falling observer. 
We may observe that $u^{\mu}=e_{(0)}\,^{\mu}=(1,0,0,v_{s}f_{s})$, which is the four-velocity (29) of the so-called Eulerian observers in the $3+1$ formalism. It should be noted that these observers are at rest with respect to the spatial infinity, as $f_{s} \rightarrow 0$ when $r\rightarrow\infty$. The same coordinate system is used for the observers adapted to the tetrads (55) and for those adapted to the tetrads (35).
The torsion tensor components (\ref{eq4}) can be evaluated for the tetrads (\ref{eq54}) and the components (\ref{eq9}) obtained. Performing this, we obtain
\begin{equation}\label{eq55}
\Sigma^{'(0)0\mu}=0\Rightarrow P^{'(0)}=0\,,
\end{equation}
i.e., a free-falling observer measures zero total energy. The same result was obtained in Ref. \cite{maluf2007reference} in the context of the Schwarzschild black hole, i.e., an observer free-falling radially into the black hole does not measure the energy of the black hole. 
In the latter Ref. this was interpreted as a consequence of the equivalence principle. Although the observer measures zero total energy, the energy of the source and gravitational energy are not zero but are equal in modulus and opposites in signal. We return to the discussion of the source and gravitational energy in section \ref{conclusions}.

\subsection{Gravitational radiation measured by a free-falling observer}\label{geodesic_radiation}

\begin{figure}
	\centering
		\includegraphics[width=.7\textwidth]{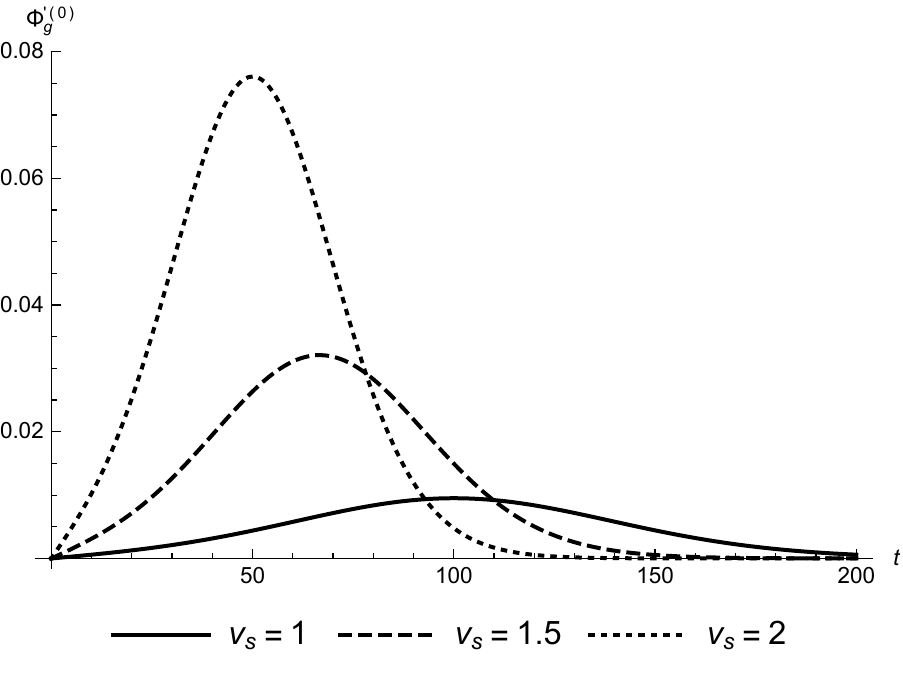}
		\caption{Gravitational flux (\ref{eq59}) measured by a free-falling observer as a function of $t$. The integrals were calculated over a cylinder of radius $\rho_{0}$ and length $2L$. The parameters are $R=1$, $\sigma=0.01$, $L=10$, $\rho_{0}=100$, and $f_{s}$ given by (\ref{eq32}).\\}
		\label{fig9}
\end{figure}

The gravitational energy flux measured by the free-falling observer can be obtained similarly to the case for the static observer. The non-zero $t^{'\mu\nu}$ components read
\begin{align}
t^{'00}=k\frac{\rho^{2}v_{s}^{2}}{2r_{s}^{2}}f^{'2}_{s}\,,\label{eq56}\\
t^{'03}=k\frac{\rho^{2}v_{s}^{3}}{2r_{s}^{2}}f_{s}f^{'2}_{s}\,.\label{eq57}
\end{align}
Thus, the non-vanishing component of $\phi_{g}^{'(0)i}$ is
\begin{equation}\label{eq58}
\phi_{g}^{'(0)3}=k\frac{\rho^{3} v_{s}^{3}}{2r_{s}^{2}}f_{s}f^{'2}_{s}\,.
\end{equation}
Hence, the gravitational energy flux measured by the free-falling observer, on the cylindrical surface, reads
\begin{equation}\label{eq59}
\Phi_{g}^{'(0)}=\frac{1}{16}\Bigg[\int_{0}^{\rho_{0}}d\rho \, \Bigg(\frac{ \rho^{3} v_{s}^{3}}{r_{s}^{2}}f_{s}f^{'2}_{s}\Bigg)\Bigg|_{z=L}-\int_{0}^{\rho_{0}}d\rho \, \Bigg(\frac{ \rho^{3} v_{s}^{3}}{r_{s}^{2}}f_{s}f^{'2}_{s}\Bigg)\Bigg|_{z=-L}\Bigg]\,.
\end{equation}
The results for the numerical integration of (\ref{eq59}) can be seen in Figure \ref{fig9}.
The gravitational energy flux decays over time as the bubble moves away from the surface. In Figure \ref{fig9}, the flux increases and then decreases. This happens when the bubble approaches the surface of the cylinder ahead and the flux increases; and as it moves away, the flux decreases. 
The larger the velocity $v_{s}$, the sooner the surface is reached by the bubble. Remember that the coordinate system is centered at Earth, so the surface is fixed with respect to Earth and moves with respect to the bubble. We can see that the integrals (\ref{eq59}) are no longer irregular when $v_{s}>1$, as was the case with (\ref{eq49}). An observer who moves with the bubble has no causality problems when measuring physical properties of the bubble.

It is interesting for a co-moving observer to measure the radiation from a source. This is not a unique feature of gravitation. It was found in Ref. \cite{maluf2010electrodynamics} that an observer moving along an accelerated charge measures electromagnetic radiation. 

\section{The variable velocity case}\label{vt}

So far, we have only considered the case $v_{s}=constant$. In this case, the energy of the source and the gravitational energy are constant in time.
In a realistic situation, the ship, hence the bubble, must accelerate from zero velocity to an arbitrary desired velocity. In the original article \cite{alcubierre1994warp}, Alcubierre considered the velocity to be a time-dependent quantity. Considering $v_{s}=v_{s}(t)$ in the metric tensor, we obtain the energy density of the source
\begin{equation}\label{eq60}
T^{00}=-\frac{1}{32\pi}\rho^{2}v_{s}(t)^{2} \left(\frac{f^{'}_{s}}{r_{s}}\right)^{2}\,,
\end{equation}
which is a function of $t$. As pointed out in \cite{bobrick2021introducing} this implies a violation of energy conservation for the source, since its energy must increase with time in an accelerated motion.

In this section, we pursuit the main objective of this article, i.e., to demonstrate that \textit{when the full gravitational system is considered, source plus gravitational field, there is no violation of energy conservation}. 
Considering $v_{s}=v_{s}(t)$ in the tetrads (\ref{eq34}), we may compute the torsion tensor components and the total, gravitational and source energies, as in section \ref{static}. By doing it, we analytically obtain the same expressions (\ref{eq42}), (\ref{eq45}) and (\ref{eq46}), respectively, but now with $v_{s}=v_{s}(t)$, i.e., a non-constant global velocity. Therefore, as in the case $v_{s}=constant$, we have $P^{(0)}=0$ when the surface integral is evaluated at infinity. Thus, as $P^{(0)}$ is not a function of time, there is no (total) energy violation when the velocity is a function of time. The total energy is still conserved in an accelerated motion. 

\begin{figure}
\centering
\begin{minipage}{.47\textwidth}
\centering
\includegraphics[width=1\textwidth]{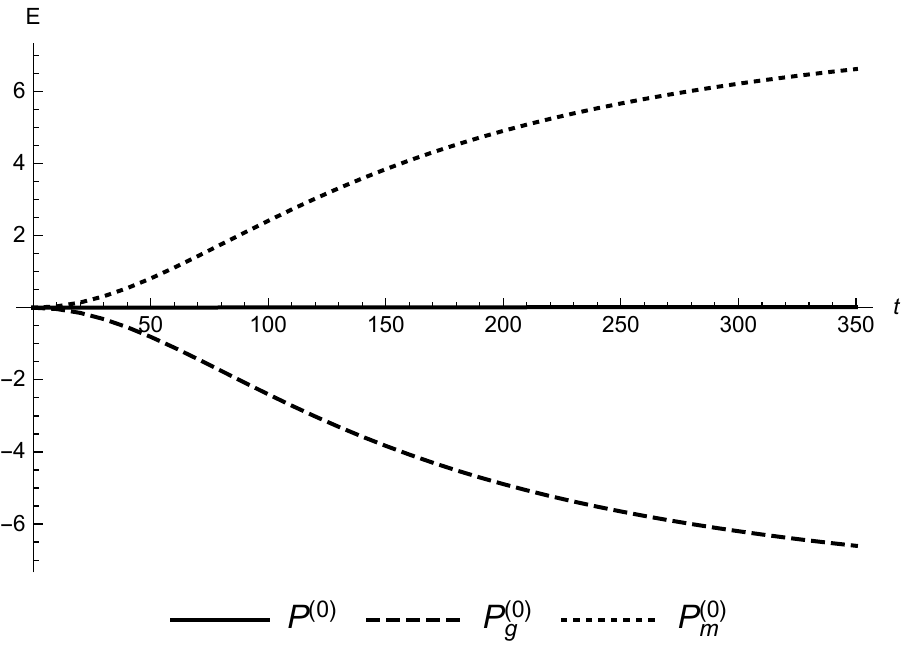}
\caption{Total (\ref{eq42}), gravitational (\ref{eq45}) and source (\ref{eq46}) energies $E$ measured by a static observer as a function of the time $t$. The integrals were calculated over a infinite cylinder of radius $\rho_{0}$. The parameters are $R=10$, $\sigma=0.01$, $\rho_{0}=300$, $\alpha=0.01$, with $f_{s}$ given by (\ref{eq32}) and $v_{s}$ by (\ref{anexa1}).}
\label{fig10}
\end{minipage}
\qquad
\begin{minipage}{.47\textwidth}
\centering
\includegraphics[width=1\textwidth]{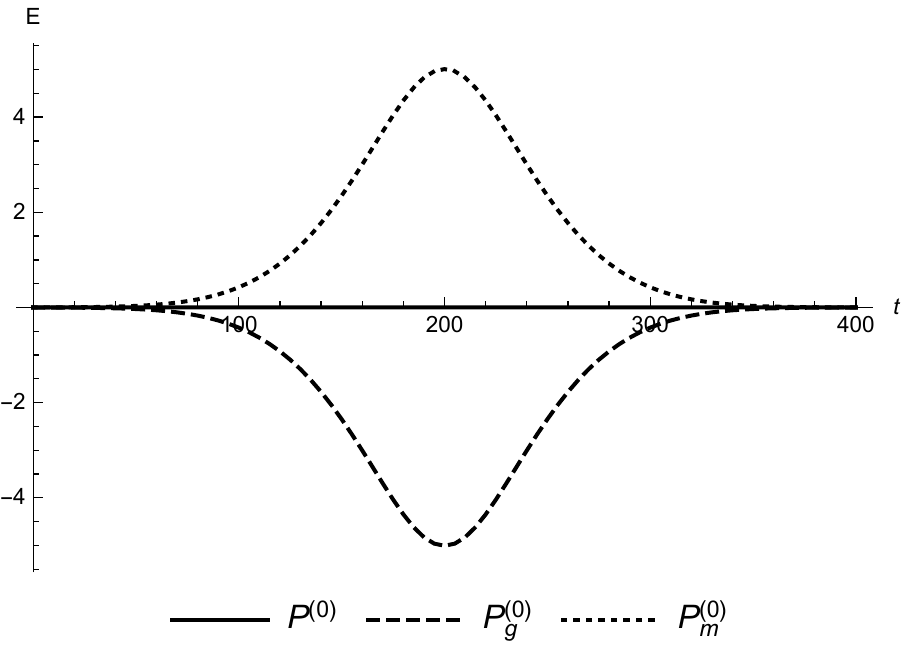}
\caption{Total (\ref{eq42}), gravitational (\ref{eq45}) and source (\ref{eq46}) energies $E$ measured by a static observer as a function of the time $t$. The integrals were calculated over a infinite cylinder of radius $\rho_{0}$. The parameters are $R=10$, $\sigma=0.01$, $\rho_{0}=400$, $v_{m}=0.9$, $a=100$, $t_{m}=200$, with $f_{s}$ given by (\ref{eq32}) and $v_{s}$ by (\ref{anexa2}).}
\label{fig11}
\end{minipage}
\end{figure}

Physically, the source energy (positive) varies with the same ``frequency'' as the gravitational energy (negative), exactly canceling out the variations of each other. In order to better elucidate what is happening, let us consider two particular velocities $v_{s}(t)$. First, the velocity of a particle with constant proper acceleration in special relativity
\begin{equation}\label{anexa1}
v_{s} = \frac{\alpha t}{\sqrt{1+\alpha^{2}t^{2}}}\,,
\end{equation}
where $\alpha$ is its proper acceleration. In Figure \ref{fig10} the results for the total, gravitational, and source energies can be seen in a region composed of an infinite cylinder with a very large radius (approximately at infinity). The variation of the gravitational energy is exactly the same as that of the source, but with the opposite sign. 
In the case $v_{s}=constant$, we have straight lines. The curve of total energy $P^{(0)}$ is zero, thus coinciding with the $t$ axis. Second, we consider a bubble with a Gaussian acceleration, i.e., which, starting from rest, accelerates to a maximum value and then decelerates to rest again. Such physical situation can be described by \cite{bobrick2021introducing}
\begin{equation}\label{anexa2}
v_{s} = v_{m}e^{-\left(\frac{t-t_{m}}{T}\right)^{2}}\,,
\end{equation}
where $v_{m}$ is the maximum value for the velocity, $t_{m}$ the instant at which the maximum velocity is attained and $2T$ is related to the time length of the trajectory. The energies for (\ref{anexa2}) can be seen in Figure \ref{fig11}. Again, the time variation of the gravitational energy plus source energy is zero.

\section{Final considerations}\label{conclusions}

Much attention has been devoted to the Alcubierre class of spacetimes, sometimes called ``warp drives''. Several related physical issues have been highlighted regarding these solutions and a vast amount of papers have attempted to address these problems. Many of these articles that have appeared recently dealt with the problem of violation of the energy conditions. Although the energy conditions are not a physical law, there is some intuition on the attractive nature of gravity, which may be violated, in a spacetime without rotation, according to the Raychaudhuri equation in the absence of the energy conditions \cite{poisson2004relativist}. Nevertheless, a violation of the strong energy condition has already been speculated for the cosmological fluid \cite{riess2004type}. In addition, negative energy density has already been predicted for the Casimir effect \cite{sopova2002energy,ulhoa2019gravitational}. Several authors have proposed Alcubierre spacetime models that do not violate some energy conditions \cite{lentz2021breaking,fell2021positive}, and Bobrick introduced an algorithm to construct such spacetimes \cite{bobrick2021introducing}. Although it has been argued that these generalizations do not solve the problem, as only one type of observer is considered \cite{santiago2021generic}. Some of these problems are unique to GR, as some of them do not seem to appear in modified theories of gravity, such as conformal gravity \cite{varieschi2013conformal}. Indeed, there are many currently unexplained features of the Alcubierre spacetime, but this spacetime is a valid class of solutions to Einstein equations, and its comprehension may lead to the solution of many, if not all, of these problems.

In the present paper, we addressed the problem of energy conservation of the spacetime, but we also uncovered new properties of the spacetime, in particular the gravitational energy flux measured by a static observer (\ref{eq49}) and by a free-falling observer (\ref{eq59}). As far as we know, the total energy of Alcubierre spacetime has never been obtained.
The presence of negative energy density is recurrent in the presence of supraluminal effects \cite{olum1998superluminal}, but we found positive energy for the source measured by a static observer.
The problem of negative energy may be a problem of the frame, i.e., of the motion of an observer interfering with the measurement of the energy.
Evaluating the gravitational energy density $e' e^{'(0)}\,_{\mu}t'^{0\mu}$ for the tetrads (\ref{eq54}), we obtain
\begin{equation}\label{eq61}
P_{g}^{'(0)}=\frac{k}{2}\int d^{3}x \, \Bigg( \rho^{3} \frac{v_{s}^{2}}{r_{s}^{2}}f^{'2}_{s} \Bigg) \geq 0\,,
\end{equation}
where $P_{g}'^{(0)}$ is the gravitational energy measured by the free-falling observer.
For the source, we have
\begin{equation}\label{eq62}
P_{m}^{'(0)}=-2k\int d^{3}x \, e^{'}e^{'}_{(0)0}G^{00}=-\frac{k}{2}\int d^{3}x \, \Bigg( \rho^{3} \frac{v_{s}^{2}}{r_{s}^{2}}f^{'2}_{s} \Bigg) \leq 0\,,
\end{equation}
where $P_{m}'^{(0)}$ is the energy of the source measured by the free-falling observer.
Thus, a free-falling observer does measure negative energy for the source, but a static does not. We can see that (\ref{eq61}) and (\ref{eq62}) satisfy (\ref{eq55}), i.e.,
\begin{equation}\label{eq63}
P_{g}^{'(0)}=-P_{m}'^{(0)}\,.
\end{equation} 
This comparison is only possible when we treat the static observer and free-falling one in the same coordinate system, as the tetrads (\ref{eq34}) and (\ref{eq54}) represent the different observers in the same coordinate system. When an observer is measuring physical properties, its dynamics affect the measurement, as is the case of a free-falling observer in the Schwarzschild spacetime, which does not measure the energy of the black hole due to its dynamic state \cite{maluf2007reference}. Therefore, only static observers can measure reliable physical properties of spacetimes. The dependence of the energy measurement on the choice of the observer is a characteristic of the definition of the energy-momentum vector itself.  This situation is not exclusive to the expression obtained in the TEGR, it also exists in special relativity.  Thus, energy measurements tell us both the information about the physical configuration and the reference system adopted.

When considering the effect of the gravitational field, in the realm of TEGR, we found that the energy of the source is dominant around the bubble, but as we radially move, the gravitational energy becomes relevant and, because it has an opposite signal, cancels out exactly the source energy, thus resulting in vanishing total energy at space infinity, as can be perceived in Figure \ref{fig1}. 
This is true even when the velocity of the bubble is a function of $t$, as demonstrated in section \ref{vt}. This is a very interesting behavior, as the Alcubierre spacetime is a spacetime with zero total energy. The non-uniformity of the source (dominant closer to the bubble) and gravitational (dominant furthest from the bubble) energies produces the desired effect of the motion for the bubble.
If we interpret the total energy at infinity as the irreducible mass of the system, as is true in Schwarzschild spacetime, the irreducible mass of Alcubierre spacetime is zero. 
Therefore, the bubble as a whole (source and gravitational field) acts as a system of null mass accelerated to supraluminal velocities.
The presence of gravitational energy flux corroborates the idea that the Alcubierre spacetime is an entity composed of a local source and a local gravitational field.
The consideration of only the source renders the analysis incomplete. The presence of gravitational energy flux may be of practical use. It is possible for observers, static and co-moving, to measure this gravitational energy flux at great distances from the bubble, thus making it possible, at least theoretically, to detect such spacetimes at a distance. 
This spacetime has a local ``warp trail'' characteristic that may even allow the direction of movement to be determined. If we make $t\rightarrow-t$ in Figure \ref{fig7}, a bubble approaching the observer rather than moving away, we obtain a reflected behavior in the $\rho_{0}$ axis, i.e., in the regions where the flux is positive in Figure \ref{fig7} it becomes negative and vice-versa.

If, on the one hand, the relativity of velocity is an old and well-founded idea, even experimentally, on the other hand, there is controversy regarding the application of this concept to acceleration.  Regarding electromagnetic radiation, we can mention the following result obtained in Ref. \cite{annalen}. An accelerated charge radiates when measured by a stationary observer, but an accelerated observer does not observe electromagnetic radiation coming from a particle at rest.  This asymmetrical situation calls into question the relativity of acceleration.  In this article, we also look at an interesting case involving acceleration. The flux of gravitational energy measured by a stationary observer remains negative for most of the temporal evolution, while for an accelerated observer the flux is positive. The total energy conservation, that is, when gravitational energy is considered, occurs when the spacecraft is accelerated when measured by a stationary observer. This feature potentially reveals yet another hint of the absolute character of acceleration.

In this article, we also were able to corroborate the extremely high energy values associated with the Alcubierre spacetime. Let us consider a realistic case for the parameters. Considering $R=100$ and $\sigma=0.1$, we obtain $P^{(0)}_{m}=-P^{(0)}_{g} \approx 8.32$ in geometrized units. This corresponds to approximately $10^{45}J$ in SI units. 
This is a relevant percentage of the rest mass of the sun ($\approx 10^{47}\,J$). High values for the energies involved in Alcubierre spacetime were also found when applying quantum inequality restriction to the Alcubierre spacetime \cite{pfenning1997unphysical}. However, the values for the energies obtained in the present article, even been beyond practical considerations, are not physically unattainable. Besides, the metric can be modified in order to reduce the energy requirements \cite{van1999awarp}. 
An analysis of the energies of the modified Alcubierre spacetimes is an interesting way to compare these spacetimes. This will be pursued elsewhere.

\section*{Acknowledgements}

The author F. L. Carneiro is grateful to L. S. Barbosa for bringing this matter to his attention.

\end{document}